\documentclass[12pt]{article}
\usepackage{color}
\usepackage{amssymb,latexsym,bm, amsmath,amsthm, amsfonts,multirow, graphicx,mathrsfs}
\usepackage{eufrak,booktabs}
\usepackage{epsfig}
\usepackage{enumitem}
\graphicspath{{figure/}}

\newcommand{\argmin}{\mathop{\rm argmin}}
\newcommand{\argmax}{\mathop{\rm argmax}}

\include{bold}

\def\vec{{\rm vec}}

\def\vecp{{\rm vecp}}

\def\1{{\bm 1}}
\def\0{{\bm 0}}

\def\cov{{\rm cov}}

\def\det{{\rm det}}
\def\tr{{\rm tr}}

\def\tparallel{{\!\!/\!/}}

\def\empirical{{\widehat{g}_n}}
\newcommand{\real}{\mathbb{R}}
\renewcommand{\tilde}{\widetilde}
\renewcommand{\hat}{\widehat}

 \def\M{{\cal M}}  
 
\def\O{{\cal SO}}

\newtheorem{prop}{Proposition}
\newtheorem{thm}{Theorem}

\newtheorem{rmk}[thm]{Remark}
\newtheorem{cor}[thm]{Corollary}

\begin{document}

\title{Robust Independent Component Analysis\\ via Minimum Divergence Estimation\\[2ex]}
\author{Peng-Wen Chen$^a$, Hung Hung$^b$\footnote{Corresponding author. E-mail: \texttt{hhung@ntu.edu.tw}.},\, Osamu Komori$^c$\\
[1ex] Su-Yun Huang$^d$, and Shinto Eguchi$^e$\\[5ex]
$^a$Department of Mathematics, National Taiwan University, Taiwan\\[1ex]
$^b$Institute of Epidemiology and Preventive Medicine\\[-1ex] National Taiwan University, Taiwan\\[1ex]
$^c$School of Statistical Thinking\\[-1ex] Institute of Statistical Mathematics, Japan\\[1ex]
$^d$Institute of Statistical Science, Academia Sinica, Taiwan\\[1ex]
$^e$Institute of Statistical Mathematics, Japan\\[5ex]}
\date{September, 2012}
\maketitle

\clearpage
\begin{abstract}
Independent component analysis (ICA) has been shown to be useful in
many applications. However, most ICA methods are sensitive to data
contamination and outliers. In this article we introduce a general
minimum $U$-divergence framework for ICA, which covers some standard
ICA methods as special cases. Within the $U$-family we further focus
on the $\gamma$-divergence due to its desirable property of super
robustness, which gives the proposed method $\gamma$-ICA.
Statistical properties and technical conditions for the consistency
of $\gamma$-ICA are rigorously studied. In the limiting case, it
leads to a necessary and sufficient condition for the consistency of
MLE-ICA. This necessary and sufficient condition is weaker than the
condition known in the literature. Since the parameter of interest
in ICA is an orthogonal matrix, a geometrical algorithm based on
gradient flows on special orthogonal group is introduced to
implement $\gamma$-ICA. Furthermore, a data-driven selection for the
$\gamma$ value, which is critical to the achievement of
$\gamma$-ICA, is developed. The performance, especially the
robustness, of $\gamma$-ICA in comparison with standard ICA methods
is demonstrated through experimental studies using simulated data
and image data.\\

\end{abstract}

\clearpage

\section{Introduction}\label{sec.1}

Consider the following generative model for independent component
analysis (ICA)
\begin{eqnarray}
X=A S+\mu,\label{ICA}
\end{eqnarray}
where the elements of the non-Gaussian source vector $S\in
\mathbb{R}^p$ are mutually independent with zero mean, $A\in
\mathbb{R}^{p\times p}$ is an unknown nonsingular mixing matrix,
$X\in \mathbb{R}^p$  is an observable random vector (signal), and
$\mu=E(X)\in \mathbb{R}^p$ is a shift parameter. Let
$Z=\Sigma^{-1/2}(X-\mu)$ be the whitened data of $X$, where
$\Sigma=\cov(X)$. An equivalent expression of model~(\ref{ICA}) in
$Z$-scale is
\begin{eqnarray}
Z=\tilde A S,\label{ICA.z}
\end{eqnarray}
where $\tilde A =\Sigma^{-1/2}A$ is the mixing matrix in $Z$-scale.
It is reported in literature that prewhitening the data can make the
ICA inference procedure more stable. In the rest of the discussion,
we will work with model~(\ref{ICA.z}) in estimating the mixing
matrix $\tilde A$ based on the prewhitened $Z$. It is easy to
transform back to the original $X$-scale via $A=\Sigma^{1/2}\tilde
A$. Note that both $\tilde A$ and $S$ are unknown, and there exists
the identifiability problem. This can be seen from the fact that
$Z=\tilde A S=(\tilde A M)(M^{-1}S)$ for any nonsingular diagonal
matrix $M$. To make $\widetilde A$ identifiable, we assume the
following conventional conditions for $S$:
\begin{eqnarray}
E(S)=0 \quad{\rm and}\quad\cov(S)=I_p,\label{assumption}
\end{eqnarray}
where $I_p\in \mathbb{R}^{p\times p}$ is the identity matrix. It
then implies that $\Sigma=AA^\top $ and
\begin{eqnarray}
I_p=\cov(Z)=\tilde A\,\cov(S)\,\tilde A^\top =\tilde A\tilde A^\top
,\label{z_constraint}
\end{eqnarray}
which means that the mixing matrix $\tilde A$ in $Z$-scale is
orthogonal. We will use notation ${\cal O}_p$ to denote the space of
orthogonal matrices in $\mathbb{R}^{p\times p}$. Note that, if
$\tilde A\in{\cal O}_p$ is a parameter of model~(\ref{ICA.z}), so is
$-\tilde A\in{\cal O}_p$. Thus, to fix one direction, we consider
$\tilde A\in \O_p$, where $\O_p\subset {\cal O}_p$ consists of
orthogonal matrices with determinant one. This set $\O_p$ is called
the special orthogonal group. The main purpose of ICA is to estimate
the orthogonal $\tilde A\in\O_p$ based on the whitened data
$\{z_i\}_{i=1}^n$, or equivalently, to look for a recovering matrix
$W\in\O_p$ so that components in $Y=:W^\top Z=(w_1^\top
Z,\ldots,w_p^\top Z)^\top$ have the maximum degree of independence.
In the latter case, $W$ provides an estimate of $\widetilde A$.

We first briefly review some existing methods for ICA. One
idea is to estimate $W$ via \textit{minimizing the mutual
information}. Let $g_Y$ be the joint probability density function of $Y=(Y_1,\dots,Y_p)^\top$, and $g_{Y_j}$ be
the marginal probability density function of $Y_j$. The mutual information, denoted by
$I(Y_1,\dots,Y_p)$, among random variables $(Y_1,\ldots,Y_p)$, is
defined to be
\begin{eqnarray}
I(Y_1,\ldots,Y_p):=\sum_{j=1}^pH(Y_j)-H(Y),\label{MI}
\end{eqnarray}
where $H(Y)=-\int g_Y\ln g_Y$ and $H(Y_j)=-\int g_{Y_j}\ln g_{Y_j}$
are the Shannon entropy. Ideally, if $W$ is properly chosen so that
$Y$ has independent components, then $g_Y=\prod_j g_{Y_j}$ and, hence,
$I(Y_1,\ldots,Y_p)=0$. Thus, via minimizing $I(Y_1,\ldots,Y_p)$ with
respect to $W$, it leads to an estimate of $W$. Another method is to estimate
$W$ via \textit{maximizing the negentropy}, which is equivalent to
minimizing mutual information as described below. The negentropy of
$Y$ is defined to be
\begin{eqnarray}
J(Y)=H(Y')-H(Y),
\end{eqnarray}
where $Y'$ is a Gaussian random vector having the same covariance
matrix as $Y$ (Hyv\"arinen and Oja, 2000). It can be deduced that
\begin{eqnarray}
I(Y_1,\ldots,Y_p)=J(Y)-\sum_{j=1}^p J(Y_j)-H(Y')+\sum_{j=1}^p
H(Y_j')=J(Y)-\sum_{j=1}^p J(Y_j),
\end{eqnarray}
where the second equality holds since, by $\cov(Y')=\cov(Y)=I_p$,
$H(Y')=\sum_{j=1}^p H(Y_j')$. Moreover, as $Y=W^\top Z$ with
$W\in\O_p$, we have $J(Y)=J(Z)$, which does not depend on $W$. That
is, the negentropy is invariant under orthogonal transformation.
Thus, minimizing the mutual information $I(Y_1,\ldots,Y_p)$ is
equivalent to maximizing the negentropy $\sum_{j=1}^p J(Y_j)$. The
negentropy $J(Y_j)$, however, involves the unknown density
$g_{Y_j}$. To avoid nonparametric estimation of $g_{Y_j}$, one can
use the following approximation (Hyv\"arinen, 1998) via a
non-quadratic contrast function $G(\cdot)$,
\begin{eqnarray}
J(Y_j)\approx J_G(Y_j)=[E\{G(Y_j)\}-E\{G(\nu)\}]^2,\label{NE_approx}
\end{eqnarray}
where $\nu$ is a random variable having the standard normal distribution. Here $J_G$ can be
treated as a measure of non-Gaussianity, and minimizing the sample
analogue of $J_G(Y_j)$ to search $W$ corresponds to the fast-ICA
(Hyv\"arinen, 1999).

Another widely used estimation criterion for $W$ is via
\textit{maximizing the likelihood}. Under model~(\ref{ICA.z}) and by
modeling $g_{Y_j}=f_j$ with some known probability density function $f_j$, the
density function of $Z$ takes the form
\begin{eqnarray}
f_Z(z;W)&=&|\det(W)|\, \prod_{j=1}^pf_j(w_j^\top z)=
\prod_{j=1}^pf_j(w_j^\top z)\label{likelihood.z}
\end{eqnarray}
since $W\in\O_p$ and hence $\det(W)=1$. The MLE-ICA then searches
the optimum $W$ via
\begin{eqnarray}
\argmin_{W\in \O_p}\mathcal{D}_0\left(\empirical,
f_Z(\cdot\,;W)\right),\label{mle_ICA_z}
\end{eqnarray}
where $\mathcal{D}_0(g,f)=-\int g\ln(f/g)$ is the Kullback-Leibler
divergence (KL-divergence), and $\empirical$ is the empirical
distribution of $\{z_i\}_{i=1}^n$. Possible choices of $f_j$ include
$f_j(s)=c_1\exp(-c_2s^4)$ for sub-Gaussian models, and
$f_j(s)=c_1/{\rm cosh}(c_2s)$ for super-Gaussian models, where
$c_1$ and $c_2$ are constants so that $f_j$ is a probability density function. It can be seen
from (\ref{likelihood.z}) that, for any row permutation matrix
$\Pi$, we have $f_Z(z;\Pi\, W)=f_Z(z; W)$. That is, we can estimate
and identify $\tilde A$ only up to its row-permutation.

As will become clear later that the above mentioned methods are all
related to \textit{minimizing the KL-divergence}, which is not
robust in the presence of outliers. Outliers, however, frequently
appear in real data analysis, and a robust ICA inference procedure
becomes necessary. For the purpose of robustness, instead of the
KL-divergence, Mihoko and Eguchi (2002) considers the
\textit{minimum $\beta$-divergence} estimation for $W$
($\beta$-ICA). The issues of consistency and robustness of
$\beta$-ICA are discussed therein. On the other hand, the
$\gamma$-divergence, which can be induced from $\beta$-divergence,
is shown to be super robust (Fujisawa and Eguchi, 2008) against data
contamination. It is our aim in this paper to propose a unified ICA
inference procedure by minimum divergence estimation. Moreover, due
to the property of super robustness, we will focus on the case of
$\gamma$-divergence and propose a robust ICA procedure, called
$\gamma$-ICA. Hyv\"arinen, Karhnen and Oja (2001) have provided a
sufficient condition to ensure the validity of MLE-ICA under the
orthogonality constraint of $W$, in the sense of being able to
recover all independent components. Amari, Chen, and Cichocki (1997)
studied necessary and sufficient conditions for consistency under a
different constraint of $W$, and this consistency result is further
extended by Mihoko and Eguchi (2002) to the case of $\beta$-ICA. In
this work, we also derive necessary and sufficient conditions for
the consistency of $\gamma$-ICA. In the limiting case $\gamma\to0$,
our necessary and sufficient condition for the consistency of
MLE-ICA is weaker than the condition stated in Hyv\"arinen, Karhnen
and Oja (2001). To the best of our knowledge, this result is not
explored in existing literature.

Some notation is defined here for the convenience of reference. For
any $M\in \mathbb{R}^{p\times p}$, let $K_p\in \mathbb{R}^{p^2\times
p^2}$ be the commutation matrix such that $\vec(M^\top)=K_p\vec(M)$;
$M>0$ (resp. $<0$) means $M$ is strictly positive (resp. negative)
definite; and $\exp(M):=\sum_{k=0}^\infty \frac{M^k}{k!}$ is the
matrix exponential. Note that $\det(\exp(M))=\exp(\tr(M))$ for any
nonsingular square matrix $M$. For a lower triangular matrix $M$
with 0 diagonals, $\vecp(M)$ stacks the nonzero elements of the
columns of $M$ into a vector with length $p(p-1)/2$. There exist
matrices $P\in \mathbb{R}^{p(p-1)/2\times p^2}$ and $Q \in
\mathbb{R}^{p^2\times p(p-1)/2}$ such that $\vecp(M)=P\vec(M)$ and
$\vec(M)=Q\vecp(M)$. Each column vector of $Q$ is of the form
$(e_i\otimes e_j)$, $i<j$, where $e_i\in\mathbb{R}^p$ is a vector
with a one in the $i$-th position and 0 elsewhere, and $\otimes$ is
the Kronecker product. $I_p\in \mathbb{R}^{p\times p}$ is the
identity matrix and $1_p\in \mathbb{R}^{p}$ is the $p$-vector of
ones.

The rest of this paper is organized as follows. A
unified framework for ICA estimation by minimum divergence is introduced in Section~2.
A robust $\gamma$-ICA procedure is developed in Section~3, wherein
the related statistical properties are studied. A geometrical
implementation algorithm for $\gamma$-ICA is further illustrated in
Section~4. In Section~5, the issue of selecting $\gamma$ value is
discussed. Numerical studies are conducted in Section~6 to
demonstrate the robustness of $\gamma$-ICA. The paper is ended with
a conclusion in Section~7. All the proofs are placed in Appendix.

\section{Minimum ${\bm U}$-divergence estimation for ICA}

In this section we introduce a general framework for ICA by means of
a minimum $U$-divergence, which covers the existing methods reviewed
in Section~1. The aim of ICA is to search a matrix $W\in\O_p$ so
that the joint probability density function $g_Y$ for $Y=W^\top Z$
is as close to marginal product $\prod_j g_{Y_j}$ as possible. This
aim then motivates estimating $W$ by minimizing a distance metric
between $g_Y$ and $\prod_j g_{Y_j}$. A general estimation scheme for
$W$ can be formulated through the following minimization problem
\begin{eqnarray}
\min_{W\in\O_p}\mathcal{D}(g_Y,\prod_j g_{Y_j}), \label{div_ica}
\end{eqnarray}
where $\mathcal{D}(\cdot,\cdot)$ is a divergence function. Different choices
of $\mathcal{D}$ will lead to different estimation criteria for ICA. Here we
will consider a general class of divergence functions, the
$U$-divergence (Murata et al., 2004; Eguchi, 2009), as described below.

The $U$-divergence is a very general class of divergence functions.
Consider a strictly convex function $U(t)$ defined on $\real$, or on
some interval of $\real$ where $U(t)$ is well-defined. Let
$\xi=\dot U^{-1}$ be the inverse function of $\dot U:=\frac{d}{dt}
U(t)$. Consider
\begin{eqnarray}
\mathcal{D}_U(g,f)&=& \int U(\xi(f))-U(\xi(g))- \dot U(\xi(g))\cdot\{\xi(f)-\xi(g)\}\nonumber\\
&=& \int U(\xi(f))-U(\xi(g))- g\cdot\{\xi(f)-\xi(g)\},
\end{eqnarray}
which defines a mapping from $\M_U\times\M_U$ to $[0,\infty)$, where
$\M_U=\left\{f: \int U(\xi(f))<\infty, \; f\ge 0 \right\}$. Define
the $U$-cross entropy by
\begin{eqnarray}
C_U(g,f)=-\int \xi(f)g +\int U(\xi(f)),
\end{eqnarray}
and the $U$-entropy by $H_U(g)=C_U(g,g)$. Then the $U$-divergence
can be written as
\begin{eqnarray}
\mathcal{D}_U(g,f)=C_U(g,f)-H_U(g)\ge0.\label{D_U}
\end{eqnarray}
In the subsequent subsections, we will introduce some special cases of
$U$-divergence, which will lead to specific methods of ICA.

\subsection{KL-divergence}

By taking the $(U,\xi)$ pair
\begin{equation}
U(t)=\exp(t),\quad\xi(t)=\ln t, \label{U_mle}
\end{equation}
the corresponding $U$-divergence is equivalent to the KL-divergence
$D_0$. In this case, it can be deduced that
\begin{equation}
\mathcal{D}_0(g_Y,\prod_jg_{Y_j})=I(Y_1,\ldots,Y_p),
\end{equation}
where $I(Y_1,\ldots,Y_p)$ is the mutual information defined in
(\ref{MI}). As described in Section~1 that
\begin{eqnarray}
\argmin_{W\in\O_p}I(Y_1,\ldots,Y_p)=
\argmax_{W\in\O_p}\sum_{j=1}^pJ(Y_j)\approx\argmax_{W\in\O_p}\sum_{j=1}^pJ_G(Y_j),\label{MI_NE}
\end{eqnarray}
we conclude that the following criteria, minimum mutual information, maximum negentropy, and fast-ICA, are all special cases of (\ref{div_ica}). On the other
hand, observe that
\begin{eqnarray}
\mathcal{D}_0(g_Y(y),\prod_jg_{Y_j}(y_j))=\mathcal{D}_0(g_Z(z),\prod_jg_{Y_j}(w_j^\top
z)),\label{mutual_mle}
\end{eqnarray}
where $g_Z$ is the joint probability density function of $Z$. If we consider the model
$g_{Y_j}=f_{j}$, and if we estimate $g_Z$ by its empirical
probability mass function $\empirical$, minimizing
(\ref{mutual_mle}) is equivalent to MLE-ICA in (\ref{mle_ICA_z}). In
summary, choosing the KL-divergence $D_0$ covers minimum mutual
information, maximum negentropy, fast-ICA, and MLE-ICA.

\subsection{$\bm\beta$-divergence}

Consider the convex set $\M_{\beta+1}:=\left\{f:  \int f^{\beta+1}
<\infty, ~f\ge 0\right\}$. Take the $(U,\xi)$ pair
\begin{align}
U(t) = \frac{1}{1+\beta}(1+\beta
t)^{\frac{\beta+1}{\beta}},\quad\xi(t)=\frac1\beta (t^\beta-1).
\end{align}
The resulting $U$-divergence defined on
$\M_{\beta+1}\times\M_{\beta+1}$ is calculated to be
\begin{eqnarray}
\mathcal{B}_\beta(g,f)=\frac{1}{\beta}\int(g^\beta-f^\beta)g-\frac{1}{\beta+1}\int(g^{\beta+1}-f^{\beta+1})\label{beta_div}\
\end{eqnarray}
which is called $\beta$-divergence (Mihoko and Eguchi, 2002), or
density power divergence (Basu et al., 1998). Note that
$\mathcal{B}_\beta(g,f)=0$ if and only if $f=\lambda g$ for some
$\lambda>0$. In the limiting case $\lim_{\beta\to
0}\mathcal{B}_\beta=\mathcal{D}_0$, it gives the KL-divergence. If
we replace $\mathcal{D}_0$ in (\ref{mle_ICA_z}) by
$\mathcal{B}_\beta$, it gives the $\beta$-ICA of Mihoko and Eguchi
(2002).

\subsection{$\bm\gamma$-divergence}\label{sec.gamma.div}

The $\gamma$-divergence can be obtained from $\beta$-divergence
through a $U$-volume normalization,
\[
\mathcal{D}_{\gamma}(g,f):=\mathcal{B}_\gamma\left(\alpha(g)\cdot g,
\alpha(f)\cdot f\right),
\]
where $\mathcal{B}_\gamma$ is defined the same way as
(\ref{beta_div}) with the plug-in $\beta=\gamma$, and where
$\alpha(f)$ is some normalizing constant. Here we adopt the
following normalization, called the volume-mass-one normalization,
\begin{equation}
\int U\big(\xi(\alpha(f)\cdot f(x))\big) dx =1.
\end{equation}
It leads to $\alpha(f)=(\gamma+1)^{1/(\gamma+1)} \|f\|_{\gamma+1}^{-1}$.
Then,
\begin{eqnarray}
\mathcal{D}_{\gamma}(g,f)=\frac{\gamma+1}{\gamma} \left\{1 -\int
\left(\frac{f(x)}{\|f\|_{\gamma+1}}\right)^\gamma
\frac{g(x)}{\|g\|_{\gamma+1}}dx\right\}.\label{gamma_div}
\end{eqnarray}
It can be seen that $\gamma$-divergence is scale invariant.
Moreover, $\mathcal{D}_{\gamma}(g,f)=0$ if and only if $f=\lambda g$
for some $\lambda>0$. The $\gamma$-divergence, indexed by a power
parameter $\gamma$, is a generalization of KL-divergence. In the
limiting case $\lim_{\gamma\to 0}\mathcal{D}_\gamma=\mathcal{D}_0$,
it gives the KL-divergence. It is well known that MLE (based on
minimum KL-divergence) is not robust to outliers. On the other hand,
the minimum $\gamma$-divergence estimation is shown to be super
robust (Fujisawa and Eguchi, 2008) against data contamination.
Hence, we will adopt $\gamma$-divergence to propose our robust
$\gamma$-ICA procedure. In particular, the main idea of $\gamma$-ICA
is to replace $\mathcal{D}_0$ in (\ref{mle_ICA_z}) by
$\mathcal{D}_\gamma$. Though the idea is straightforward, there are
many issues need to be studied. Detailed inference procedure and
statistical properties of $\gamma$-ICA are discussed in Section~3.

\section{The $\bm{\gamma}$-ICA inference procedure}\label{sec.gamma_ICA_Z}

The ICA is actually a two-stage process. First, we need to whiten
the data. The whitened data are then used for the recovery of independent
sources. Since the main purpose of this study is to develop a robust
ICA inference procedure, the robustness for both data
prewhitening and independent source recovery should be
guaranteed. Here we will utilize the $\gamma$-divergence to
introduce a robust prewhitening method called $\gamma$-prewhitening,
followed by illustrating $\gamma$-ICA based on the prewhitened data.
In practice, the $\gamma$ value for $\gamma$-divergence should also
be determined. In the rest of discussion, we will assume $\gamma$ is
given, and leave the discussion of its selection to
Section~\ref{sec.select_gamma}.

\subsection{$\bm{\gamma}$-prewhitening}\label{sec.gamma_prewhiten}

Although prewhitening is always possible by a straightforward standardization
of $X$, there exists the issue of robustness of such a whitening procedure. It is
well known that empirical moment estimates of $(\mu,\Sigma)$ are
very sensitive to outliers.
In Mollah, Eguchi and Minami (2007), the authors
proposed a robust $\beta$-prewhitening procedure. In particular, let
$\xi_{\mu,\Sigma}(x)$ be the probability density function of $p$-variate normal distribution
with mean $\mu$ and covariance $\Sigma$, and let $\widehat g_X(x)$
be the empirical distribution based on data $\{x_i\}_{i=1}^n$. With a
given $\beta$, Mollah et al. (2007) proposed the following minimum $\beta$-divergence
estimators
\begin{eqnarray}
(\widehat\kappa,\widehat\mu,\widehat\Sigma)=\argmin_{\kappa,\mu,\Sigma}\mathcal{B}_\beta(\widehat
g_X,\kappa\cdot\xi_{\mu,\Sigma}),
\end{eqnarray}
and then suggested to use $(\widehat\mu,\widehat\Sigma)$ for
whitening the data.  Interestingly, $(\widehat\mu,\widehat\Sigma)$
can also be derived from the minimum $\gamma$-divergence as
\begin{eqnarray}
(\widehat\mu,\widehat\Sigma)=\argmin_{\mu,\Sigma}\mathcal{D}_\gamma(\widehat
g_X,\xi_{\mu,\Sigma}).\label{gamma.prewiten.obj}
\end{eqnarray}
At the stationarity of (\ref{gamma.prewiten.obj}), the solutions
$(\hat\mu,\hat\Sigma)$ will satisfy
\begin{eqnarray}
\widehat\mu=\frac{\sum_{i=1}^nd_i^\gamma(\widehat\mu,\widehat\Sigma)\cdot
x_i} {\sum_{i=1}^nd_i^\gamma(\widehat\mu,\widehat\Sigma)}\quad{\rm
and }\quad \hat\Sigma=
(1+\gamma)\cdot\frac{\sum_{i=1}^nd_i^\gamma(\widehat\mu,\widehat\Sigma)\cdot
(x_i-\widehat\mu)(x_i-\widehat\mu)^\top}
{\sum_{i=1}^nd_i^\gamma(\widehat\mu,\widehat\Sigma)},\label{gamma.prewiten}
\end{eqnarray}
where
$$d_i(\mu,\Sigma)=\exp\left\{-\frac{1}{2}(x_i-\mu)^\top \Sigma^{-1}(x_i-\mu)\right\}.$$
The robustness property of $(\widehat\mu,\widehat\Sigma)$ can be
found in Mollah et al. (2007). We
call the prewhitening procedure
\begin{eqnarray}
z_i = \widehat\Sigma^{-1/2}(x_i-\widehat\mu), \quad i=1,\ldots,n
\end{eqnarray}
the $\gamma$-prewhitening. The whitened data $\{z_i\}_{i=1}^n$
then enter the $\gamma$-ICA estimation procedure.

\subsection{Estimation of $\gamma$-ICA}

We are now in the position to develop our $\gamma$-ICA based on the
$\gamma$-prewhitened data $\{z_i\}_{i=1}^n$. As discussed in
Section~\ref{sec.gamma.div}, the $W$ estimator is derived from
\begin{equation}
\hat W = \argmin_{W\in \O_p}\mathcal{D}_\gamma(\empirical,
f_Z(\cdot;W))
\end{equation}
where $f_Z(z;W)=\prod_{j=1}^pf_j(w_j^\top z)$ and $f_j$ is the
working model for $g_{Y_j}$. Since $W\in\O_p$,
\begin{eqnarray*}
\int f_Z^{\gamma+1}(z;W)dz=|\det(W)|\prod_{j=1}^p\int
f_j^{\gamma+1}(y_j)dy_j=\prod_{j=1}^p\int  f_j^{\gamma+1}(y_j)dy_j,
\end{eqnarray*}
which does not involve $W$. Thus, $\hat W$ can be equivalently
obtained via
\begin{eqnarray}
\hat W=\argmax_{W\in\O_p}{\cal L}(W):=\argmax_{W\in\O_p}\,
\frac1n\sum_{i=1}^n \left\{\prod_{j=1}^p f_{j}^\gamma(w_j^\top
z_i)\right\}.\label{gamma_ICA_z0}
\end{eqnarray}
Finally, the mixing matrix $A$ is estimated by $\hat
A=\hat\Sigma^{1/2}\hat W $. Let $f(W^\top
z)=\prod_{j=1}^pf_j(w_j^\top z)$ and
\[\phi(W^\top z):=\left[\phi_1(w_1^\top z),\dots,\phi_p(w_p^\top z)\right]^\top,\;\;
{\rm where}\;\;\phi_j(y)=\frac{d}{dy}\ln f_j(y).\] We have the
following proposition.

\begin{prop}\label{prop.stationarity}
At the stationarity, the maximizer $\widehat W$ defined in
(\ref{gamma_ICA_z0}) will satisfy
\begin{eqnarray}
\frac{1}{n}\sum_{i=1}^n f^{\gamma}(\widehat W^\top z_i)
\left\{\widehat W^\top z_i\, \left[\phi (\widehat W^\top
z_i)\right]^\top-\phi(\widehat W^\top z_i)\left[\widehat W^\top z_i\right]^\top\right\}=0.
\end{eqnarray}
\end{prop}

From Proposition~\ref{prop.stationarity}, it can be easily seen the
robustness nature of $\gamma$-ICA: the stationary equation is a
weighted sum with the weight function $f^\gamma$. When $\gamma>0$,
an outlier with extreme value will contribute less to the stationary
equation. In the limiting case of $\gamma\to 0$, which corresponds
to MLE-ICA, the weight $f^\gamma$ becomes uniform and, hence, is not
robust.

\subsection{Consistency of $\gamma$-ICA}

A critical point to the likelihood-based ICA method is to specify a
working model $f_j$ for $g_{Y_j}$. A sufficient condition to ensure
the consistency of MLE-ICA can be found in Hyv\"arinen, Karhnen and
Oja (2001). Here the ICA consistency means recovery consistency.
That is, an ICA procedure is said to be recovery consistent if it is
able to recover all the independent components. Note that the
consistency of MLE-ICA does not rely on the correct specification of
working model $f_j$, but only on the positivity of
$E[\phi_j(S_j)S_j-\phi'_j(S_j)]$, $j=1,\ldots,p$. This subsection
aims to investigate the consistency of $\gamma$-ICA for
$\gamma\in\Gamma=(0,\tau]$, where $\tau>0$ is some constant. We will
deduce necessary and sufficient conditions such that $\gamma$-ICA is
recovery consistent. The main result is summarized below.

\begin{thm}\label{thm.consistency}
Assume the ICA model~(\ref{ICA.z}). Assume the existence of $\tau$
for $\Gamma=(0,\tau]$ such that
\begin{enumerate}
\item[(A)]
$E[f_j^\gamma(S_j)S_j]=0\,$ for all $\gamma\in\Gamma$ and all
$j=1,\ldots,p$.
\end{enumerate}
Then, for $\gamma\in\Gamma$, the associated $\gamma$-ICA is recovery
consistent if and only if
\begin{eqnarray}
\Psi_\gamma=Q^\top \left(I_{p^2}-K_{p}\right)
\left\{\gamma\Psi_{(1)}+\gamma^2\Psi_{(2)}\right\}
\left(I_{p^2}-K_{p}\right) Q<0,\label{hessian_tangent}
\end{eqnarray}
where $\Psi_{(1)}=\sum_{j=1}^p(e_je_j^\top\otimes U_j)-(D\otimes
I_p)$, $U_j={\rm diag}(u_{j1},\ldots,u_{jp})$,
$u_{jk}=E[f^\gamma(S)\phi_j'(S_j)S_k^2]$, $D={\rm
diag}(d_1,\ldots,d_p)$, $d_j=E[f^\gamma(S)\phi_j(S_j)S_j]$, and
$\Psi_{(2)}=E[f^\gamma(S)\{\phi(S)\phi^\top(S)\otimes SS^\top\}]$.
\end{thm}

Condition~(A) of Theorem~\ref{thm.consistency} can be treated as a
weighted version of $E(S_j)=0$. It is satisfied when $S_j$ is
symmetrically distributed about zero, and when the model probability
density function $f_j$ is an even function. We believe condition~(A)
is not restrictive and should be approximately valid in practice.
Notice that $\Psi_{(2)}>0$. Thus, for the validity of
(\ref{hessian_tangent}), we must require that $\gamma\Psi_{(1)}<0$,
and the effect of $\gamma^2\Psi_{(2)}>0$ can be exceeded by
$\gamma\Psi_{(1)}<0$. Fortunately, due to the coefficient
$\gamma^2$, when $\gamma$ is small, the effect of $\gamma\Psi_{(1)}$
will eventually outnumber the effect of $\gamma^2\Psi_{(2)}$, so
that $\Psi_\gamma<0$ can be ensured. In this situation, the negative
definiteness of $\Psi_\gamma$ mainly relies on the structure of
$\Psi_{(1)}$. Moreover, a direct calculation gives
$Q^\top(I_{p^2}-K_{p})\Psi_{(1)}(I_{p^2}-K_{p})Q$ to be a diagonal
matrix with diagonal elements $\{(u_{jk}-d_j)+(u_{kj}-d_k):j< k\}$.
We thus have the following corollary.

\begin{cor}\label{cor.consistency}
Assume the ICA model~(\ref{ICA.z}). Assume the existence of a small
enough $\tau$ for $\Gamma=(0,\tau]$ such that
\begin{enumerate}
\item[(A)]
$E[f_j^\gamma(S_j)S_j]=0$ for $\gamma\in\Gamma$, $j=1,\ldots,p$.

\item[(B)]
$E[f^\gamma(S)\{\phi_j(S_j)S_j-\phi_j'(S_j)S_k^2\}]+
E[f^\gamma(S)\{\phi_k(S_k)S_k-\phi_k'(S_k)S_j^2\}]>0$ for
$\gamma\in\Gamma$, for all pairs $(j,k)$, $j\neq k$.
\end{enumerate}
Then, for every $\gamma\in\Gamma$, the associated $\gamma$-ICA can
recover all independent components.
\end{cor}

To understand the meaning of condition~(B), we first consider an
implication of Corollary~\ref{cor.consistency} in the limiting case
of $\gamma\to 0$, which corresponds to the MLE-ICA. In this case,
condition~(A) becomes $E(S_j)=0$, which is automatically true by the
model assumption of $S$. Moreover, since $E(S_j^2)=1$, condition~(B)
becomes
\begin{eqnarray}
E[\phi_j(S_j)S_j-\phi_j'(S_j)]+
E[\phi_k(S_k)S_k-\phi_k'(S_k)]>0,\quad \mbox{for all pairs $(j,k)$, $j\neq k$}.\label{consistency.mle}
\end{eqnarray}
A sufficient condition to ensure the validity of
(\ref{consistency.mle}) is
\begin{equation}
E[\phi_j(S_j)S_j-\phi_j'(S_j)]>0,\quad\forall j,
\label{consistency.mle2}
\end{equation}
which is the same condition given in Theorem~9.1 of Hyv\"arinen,
Karhnen and Oja (2001) for the consistency of MLE-ICA. We should
note that (\ref{consistency.mle}) is a weaker condition than
(\ref{consistency.mle2}). In fact, from the proof of
Theorem~\ref{thm.consistency}, (\ref{consistency.mle}) is also a
necessary condition. One implication of (\ref{consistency.mle}) is
that, we can have at most one $f_j$ to be wrongly specified or at
most one Gaussian component involved, and MLE-ICA is still able to
recover all independent components. This can also be intuitively
understood that once we have determined $p-1$ directions in
$\mathbb{R}^p$, the last direction is automatically determined.
However, this fact cannot be observed from (\ref{consistency.mle2})
which requires all $f_j$ to be correctly specified. We summarize the
result for MLE-ICA below.

\begin{cor}
Assume the ICA model~(\ref{ICA.z}). Then, MLE-ICA is recovery
consistent if and only if $E[\phi_j(S_j)S_j-\phi_j'(S_j)]+
E[\phi_k(S_k)S_k-\phi_k'(S_k)]>0$ for all pairs $(j,k)$, $j\neq k$.
\end{cor}

Turning to the case of $\gamma$-ICA, condition~(B) of
Corollary~\ref{cor.consistency} can be treated as a weighted version
of (\ref{consistency.mle}) with the weight function $f^\gamma$.
However, one should notice that the validity of $\gamma$-ICA has
nothing to do with that of MLE-ICA, since there is no direct
relationship between condition~(B) and its limiting case
(\ref{consistency.mle}). In particular, even if
(\ref{consistency.mle}) is violated (i.e., MLE-ICA fails), with a
proper choice of $\gamma$, it is still possible that condition~(B)
holds and, hence, the  recovery consistency of $\gamma$-ICA can be
guaranteed.

\begin{rmk}
By Theorem~\ref{thm.consistency}, a valid $\gamma$-ICA procedure
must correspond to $\Psi_\gamma<0$, or equivalently, the maximum
eigenvalue of $\Psi_\gamma$, denoted by
$\lambda_{\max}(\Psi_\gamma)$, must be negative. How should one pick
a $\Gamma$-interval so that $\gamma\in\Gamma$ is legitimate in the
sense that $\lambda_{\max}(\Psi_\gamma)<0$? Our suggestion for a
rule of thumb is as follows. Let $\widehat\Psi_\gamma$ be the
empirical estimator of $\Psi_\gamma$ based on the estimated source
$\{\widehat s_i\}_{i=1}^n$, where $\widehat s_i:=\widehat W^\top
z_i$. The plot of $\{(\gamma,
\lambda_{\max}(\widehat\Psi_\gamma))\}$ then provides a guidance in
determining $\Gamma$, over which
$\lambda_{\max}(\widehat\Psi_\gamma)$ should be far away below zero.
With the $\Gamma$-interval, a further selection procedure,
introduced in Section~\ref{sec.select_gamma}, can be applied to
select an optimal $\gamma$ value from $\Gamma$. It is confirmed in
our numerical study in Section~6 that, the interval $\Gamma$, where
$\lambda_{\max}(\widehat\Psi_\gamma)<0$, is quite wide, and the
suggested rule does provide adequate choice of $\Gamma$. It also
implies that the choice of $\tau$ in Corollary~\ref{cor.consistency}
is not critical, as $\tau$ is allowed to vary in a wide range and
not limited to very small number. It is the condition~(B) that plays
the most important role to ensure the recovery consistency of
$\gamma$-ICA.
\end{rmk}

\subsection{$\beta$-ICA versus $\bm\gamma$-ICA}

By using $\beta$-divergence, Mihoko and Eguchi (2002) proposed
$\beta$-ICA to recover independent components. The objective
function of $\beta$-ICA being maximized is of the form
\begin{eqnarray}
|{\rm det}(W)|^\beta\left\{\int f^\beta(W^\top z)
g_Z(z)dz-c_\beta\right\},\label{div.beta}
\end{eqnarray}
where $c_\beta$ is a known constant. If we restrict $W\in\O_p$, then
$|{\rm det}(W)|=1$ and maximizing (\ref{div.beta}) is equivalent to
maximizing $\int f^\beta(W^\top z) g_Z(z)dx$, which has the same
form with the population objective function of $\gamma$-ICA in
(\ref{gamma_ICA_z0}). We should emphasize that Mihoko and Eguchi
(2002) considered the ICA problem under the original $X$-scale,
while the constraint $W\in\O_p$ is a consequence of prewhitening.
Without considering the constraint $W\in\O_p$, the objective
function of $\gamma$-ICA is deduced to be
\begin{eqnarray} |{\rm
det}(W)|^{\frac{\gamma}{\gamma+1}}\left\{\int f^\gamma(W^\top z)
g_Z(z)dz\right\}\label{div.gamma}
\end{eqnarray}
which is different from (\ref{div.beta}). However, (\ref{div.beta})
is similar to (\ref{div.gamma}) when $c_\beta$ is small. This fact
also confirms the observation of Mihoko and Eguchi (2002) that
setting $c_\beta=0$ does not affect the performance of $\beta$-ICA.
In summary, $\gamma$-ICA and $\beta$-ICA based on the whitened
data $Z$ are equivalent. For data $X$ in original scale, however,
$\gamma$-ICA maximizing (\ref{div.gamma}) is different from
$\beta$-ICA maximizing (\ref{div.beta}), but they will have similar
performance for small $\beta$.

\section{Gradient method for $\bm{\gamma}$-ICA on $\bm{\O_p}$}

In this section, we introduce an algorithm for estimating $W$
constrained to the special orthogonal group $\O_p$, which is a Lie
group and is endowed with a manifold structure.\footnote{
$\mathcal{G}$ is a Lie group if the group operations
$\mathcal{G}\times \mathcal{G}\to \mathcal{G}$ defined by $(x,y)\to
xy$ and $\mathcal{G}\to \mathcal{G}$ defined by $x\to x^{-1}$ are
both $C^\infty$ mappings (Boothby, 1986).}  The Lie group $\O_p$,
which is a path-connected subgroup of $\mathcal{O}_p$, consists of
all orthogonal matrices in $\mathbb{R}^{p\times p}$ with determinant
one.\footnote{The reason why we consider $\O_p$ is that
$\mathcal{O}_p$ itself is not connected. In the case that the
desired orthogonal matrix $W$ has determinant $-1$, our algorithm in
fact searches for $\Pi W\in\O_p$ for some permutation matrix $\Pi$
with $\det(\Pi)=-1$.} Recall $\mathcal{L}$ being the objective
function of $\gamma$-ICA maximization problem defined in
(\ref{gamma_ICA_z0}). A desirable algorithm is to generate an
increasing sequence $\{\mathcal{L}(W_k)\}_{k=1}^\infty $ with
$W_k\in \O_p$, such that $\{W_k\}_{k=1}^\infty$ converges to a local
maximizer $W^*$ of $\mathcal{L}$. Various approaches can be used to
generate such a sequence $\{W_k\}_{k=1}^\infty$ in $\O_p$, for
instance, geodesic flows and quasi-geodesic flows (Nishimori and
Akaho, 2005).  Here we focus on geodesic flows on  $\O_p$. In
particular, starting with the current $W_k$, the update $W_{k+1}$ is
selected from one geodesic path of $W_k$ along the steepest ascent
direction such that $\mathcal{L}(W_{k+1})>\mathcal{L}(W_k)$. In
fact, this approach has been applied to the general Stiefel manifold
(Nishimori and Akaho, 2005). Below we briefly review the idea and
then introduce our implementation algorithm for $\gamma$-ICA. We
note that the proposed algorithm is also applicable to MLE-ICA by
changing the corresponding objective function.

Let $T_{W}\O_p$ denote  the tangent space of $\O_p$ at $W$. Consider
a smooth path $W(t)$ on $\O_p$ with $W(0)=W$.  Differentiating
$W(t)^\top W(t)=I_p$ yields the tangent space at $W$
\begin{equation}
T_{W}\O_p =\left\{WV: V \in\real^{p\times p},  V^\top=- V\right\}.
\end{equation}
Clearly, $T_{I_p}\O_p$ is the set of all skew-symmetric matrices.
Each geodesic path starting from $I_p$ has an intimate relation with
the matrix exponential function. In fact, $\exp(V)\in\O_p$ if and
only if $V$ is skew-symmetric (see page 148 in Boothby, 1986;
Proposition~9.2.5. in Marsden and Ratiu, 1998). Moreover, for any
$M\in\O_p$, there exists (not unique) a skew-symmetric $V$ such that
$M=\exp(V)$. If the Killing metric (Nishimori and Akaho, 2005)
\[g_W(Y_1,Y_2):=\tr(Y_1^\top Y_2),\;\; {\rm where}~ Y_1,Y_2\in T_W\O_p,\]
is used, the geodesic path starting from $I_p$ in the direction $V$
is given by
\begin{eqnarray}
\left\{\Phi(V,t):~t\in \mathbb{R}\right\}\quad{\rm with}\quad
\Phi(V,t):=\exp(t V).\label{geodesic.Ip}
\end{eqnarray}

Since the Lie group is homogeneous, we can compute the gradient and
geodesic at $W_k\in\O_p$ by pulling them back to the identity $I_p$
and then transform back to $W_k$. In the implementation algorithm,
to ensure all the iterations lying on the manifold $\O_p$, we update
$W_{k+1}$ through
\begin{equation}\label{W.update}
W_{k+1}:=W_k\exp( t_kV_k),
\end{equation}
where the skew-symmetric matrix $V_k$ and the step size $t_k$ are
chosen properly to meet the ascending condition
$\mathcal{L}(W_{k+1})>\mathcal{L}(W_k)$. Since, from
(\ref{geodesic.Ip}), $\exp(t_kV_k)$ lies on the geodesic path of
$I_p$, then $W_{k+1}=W_k\exp(t_kV_k)$ must lie on the geodesic path
of $W_k$. Moreover, since $\det(W_{k+1})=\det(W_k)\exp(0)=1$ by
$\tr(V_k)=0$, the sequence in (\ref{W.update}) satisfies $W_k\in
\O_p$ for all $k$. The determination of the gradient direction $V_k$
and the step size $t_k$ is discussed below.

To compute the gradient and geodesic at $W_k$ by pulling them back
to $I_p$, define
\begin{eqnarray}\label{F}
\mathcal{F}_{W_k}(W):=\mathcal{L}(W_kW).
\end{eqnarray}
We then determine $W_{k+1}=W_k\exp(t_kV_k)$ from one geodesic at
$I_p$ in the direction of the projected gradient of
$\mathcal{F}_{W_k}$. Specifically, to ensure the ascending
condition, we choose each skew-symmetric $V_k$ to be
$\nabla_{\tparallel} \mathcal{F}_{W_k}$, the projected gradient of
$\mathcal{F}_{W_k}$ at $I_p$,
defined to be
\begin{eqnarray}
\label{V}\nabla_{\tparallel} \mathcal{F}_{W_k}&:=&\argmin_{V\in
T_{I_p}\O_p} \|\nabla \mathcal{F}_{W_k}-V\|,\quad\mbox{where}~~
\nabla\mathcal{F}_{W_k}:=\frac{\partial \mathcal{F}_{W_k}} {\partial
W}\Big|_{W=I_p},\nonumber\\
&=&\frac12 \left(\nabla \mathcal{F}_{W_k}-\nabla
\mathcal{F}_{W_k}^\top\right)\nonumber\\
&=&\frac{\gamma}{2n}\sum_{i=1}^n f^{\gamma}(W_k^\top z_i)
\left\{W_k^\top z_i\, \left[\phi (W_k^\top
z_i)\right]^\top-\phi(W_k^\top z_i)\left[W_k^\top
z_i\right]^\top\right\}.
\end{eqnarray}
This particular choice of $V_k$ ensures the existence of the step
size $t_k$  for the ascending condition. Note that  in the case of
$\O_p$ imposed with the Killing metric, the projected gradient
coincides with the natural gradient introduced by Amari (1998). See
also Fact~5 in Nishimori and Akaho (2005) for further details. As to
the selection of the step size $t_k$ at each iteration $k$ with
$W_k$ and $V_k=\nabla_{\tparallel} \mathcal{F}_{W_k}$,
we propose to select $t_k$ such that $W_k\exp(t_k V_k)$ is the
``first improved rotation". In particular, we consider $t_k=\alpha
\rho^{\ell_k} $ for some $\alpha>0$ and $0<\rho<1$, where $\ell_k$
is the nonnegative integer. To proceed, we search $\ell_k$ such that
\[{\cal L}\big( W_k\exp(\alpha\rho^{\ell_k}V_k)\big)> {\cal L}
\big(W_k \big),\quad{\rm where}\quad V_k=\nabla_{\tparallel}
\mathcal{F}_{W_k},
\]
and then update $W_{k+1}=W_k \exp(\alpha \rho^{\ell_k}V_k)$. In our
implementation, $\alpha=1$ and $\rho=0.5$ are used. For the
convergence issue, one can instead consider the Armijo rule for
$t_k$ (given in equation~(\ref{Armijo})). Our experiments show that
the above ``first improved rotation" rule works quite well. Lastly,
in the implementation, to  save the storage for $W_k$, we ``rotate
$\bm{Z}$ directly" instead of manipulating $W$, where $\bm{Z}$ is
the $p\times n$ data matrix whose columns are $z_i$, $i=1,\dots,n$.
That is, we use the update $\bm{Z}_k= W_k^\top \bm{Z}$. To retrieve
the matrix $W$, we simply do a matrix right division of the final
$\bm{Z}$ and the initial $\bm{Z}$. The algorithm for $\gamma$-ICA
based on
gradient ascend on $\O_p$ is summarized below.\\

\hrule
\begin{enumerate}\itemsep=0pt
\item
Initialization: $\alpha=1$, $\rho=0.5$, prewhitened  data
$\bm{Z}_1=\bm{Z}$ (${p\times n}$ matrix).
\item
For each iteration $k=1,2,3,\ldots$,
\begin{itemize}\itemsep=0pt
\item[(i)] Compute the skew-symmetric matrix $V_k$ in (\ref{V}).
\item[(ii)] For $\ell_k=0,1,2,\ldots$, if $\mathcal{F}_{W_k}(\exp(\alpha \rho^{\ell_k} V_k))>{\cal F}_{W_k}(I_p)$, then break the
loop.
\item[(iii)] Update  $\bm{Z}_{k+1}$  by $\exp(\alpha \rho^{\ell_k}V_k)^\top \bm{Z}_k$.
Check the convergence criterion. If the criterion is not met, go
back to (i).
\end{itemize}
\item Output $\widehat W=\left(\bm{Z}_1\bm{Z}_1^\top\right)^{-1}\bm{Z}_1\bm{Z}_k^\top$.
\end{enumerate}
\hrule~\\

Finally, we would like to mention the convergence issue. The
statement is similar to Proposition~1.2.1 of Bertsekas (2003).
\begin{thm}\label{thm.armijo}
Let $\mathcal{L}$ be continuously differentiable on $\O_p$, and
$\mathcal{F}$ be defined in (\ref{F}). Let $\{W_k\in \O_p\}$ be a
sequence generated by $W_{k+1}=W_k \exp( t_k V_k)$, where $V_k$ is a
projected  gradient related (see (\ref{projected}) below) and $t_k$
is a properly chosen step size by the Armijo rule: reduce the step
size $t_k=\alpha \rho^{\ell_k}$, $\ell_k=0,1,2,\ldots, $ until the
inequality holds for the first nonnegative $\ell_k$,
\begin{equation}\label{Armijo}
\mathcal{L}(W_{k+1})-\mathcal{L}(W_k)=\mathcal{F}_{W_k}(\exp(t_k V_k
))-\mathcal{F}_{W_k}(I_p)\ge \eta\, t_k\, \tr\left(\nabla_\tparallel
\mathcal{F}_{W_k}^\top V_k \right),
\end{equation}
where $0<\eta<1$ is a fixed constant. Then, every limit point $ W^*$
of $\{W_k\in \O_p\}$ is a stationary point, i.e., $\tr(\nabla
\mathcal{F}_{ W^*}^\top V)=0$ for all $V\in T_{ W^*} \O_p$, or
equivalently, $\nabla_\tparallel \mathcal{F}_{ W^*}=0$.
\end{thm}
The statement that $V_k$ is a projected gradient related corresponds
to the condition
\begin{equation}\label{projected}
\limsup_{k\to \infty}\, \tr\left(\nabla_\tparallel
\mathcal{F}_{W_k}^\top V_k\right)>0.  \end{equation} This condition
is true when $V_k$ is the projected gradient $\nabla_\tparallel
\mathcal{F}_{W_k}$ itself or some natural gradient
$M^{-1}\nabla_\tparallel \mathcal{F}_{W_k}$ (Theorem 1, Amari,
1998), where $M$ is a Riemannian metric tensor, which is positive
definite.

\section{Selection of $\gamma$}\label{sec.select_gamma}

The estimation process of $\gamma$-ICA consists of two steps:
$\gamma$-prewhitening and the geometry-based estimation for $W$, in
which the values of $\gamma$ are essential to have robust
estimators. Hence, we carefully select the value of $\gamma$ based
on the adaptive selection procedures proposed by Minami and Eguchi
(2003) and Mollah et al.~(2007). We first introduce a general idea
and then apply the idea to the selection of $\gamma$ in both
$\gamma$-prewhitening and $\gamma$-ICA. Define the measurement of
generalization performance as
\begin{equation}
C_{\gamma_0}(\gamma)=E[\mathcal{D}_{\gamma_0}(g,f_{\hat\theta_\gamma})],
\end{equation}
where $g$ is the underlying true joint probability density function
of the data, $f_\theta$ is the considered model for fitting,
$\hat\theta_\gamma:=\argmin_{\theta} \mathcal{D}_\gamma(\widehat g,
f_{\theta})$ is the minimum $\gamma$-divergence estimator of
$\theta$, and $\widehat g$ is the empirical estimate of $g$. The
$\gamma_0$ is called the anchor parameter and is fixed at
$\gamma_0=1$ throughout this paper. This value is empirically shown
to be insensitive to the resultant estimators (Minami and Eguchi,
2003). Let $\widehat C_{\gamma_0}(\gamma)$ be the sample analogue of
$C_{\gamma_0}(\gamma)$. We propose to select the value of $\gamma$
over a predefined set $\Gamma$ through
\begin{eqnarray}
\widehat\gamma=\argmin_{\gamma\in\Gamma} \widehat
C_{\gamma_0}(\gamma).
\end{eqnarray}
For $\gamma$-prewhitening, $g=g_X$ and $f_\theta=\xi_{\mu,\Sigma}$
with $\theta=(\mu,\Sigma)$. For $\gamma$-ICA, $g=g_Z$ and
$f_\theta=f_Z(\cdot;W)$ with $\theta=W$.

The above selection criterion requires the estimation of
$C_{\gamma_0}(\gamma)$. To avoid the problem of overfitting, we
apply a $K$-fold cross-validation. Let $\mathcal T$ be the
whole data, and let $K$ partitions of $\mathcal T$ be $\mathcal
T_1,\ldots,\mathcal T_K$, that is, $\mathcal T_i\bigcap \mathcal
T_j=\emptyset$ if $i\not=j$ and $\mathcal T=\cup_{i=1}^K \mathcal
T_i$. The whole selection procedure is summarized below.\\

\hrule
\begin{enumerate}
\item For $k=1,\ldots,K$,
\begin{enumerate}\itemsep=0pt
\item[(i)]
For every $\gamma\in\Gamma$, obtain
$\widehat\theta_\gamma^{(-k)}:=\argmin_{\theta}C_\gamma(\widehat
g^{(-k)},f_{\theta})$, where $\widehat g^{(-k)}$ is the empirical
estimate of $g$ based on $\mathcal T\setminus\mathcal T_k$.

\item[(ii)]
Compute the cross validation estimate $C_{\gamma_0}(\widehat
g^{(k)},f_{\hat\theta_\gamma^{(-k)}})$, where $\widehat g^{(k)}$ is
the empirical estimate of $g$ based on $\mathcal{T}_k$.
\end{enumerate}

\item
Estimate $C_{\gamma_0}(\gamma)$ by
\begin{equation}
\widehat
C_{\gamma_0}(\gamma)=\frac1K\sum_{k=1}^KC_{\gamma_0}(\widehat
g^{(k)},f_{\hat\theta_\gamma^{(-k)}})\label{cv}
\end{equation}
and obtain $\widehat\gamma=\argmin_{\gamma\in\Gamma} \widehat
C_{\gamma_0}(\gamma)$.
\end{enumerate}
\hrule

~\\

Eventually, we have two optimal values of $\gamma$:
$\widehat\gamma_{\mu,\Sigma}$ for $\gamma$-prewhitening and
$\widehat\gamma_W$ for estimation of the recovering matrix $W$.

\section{Numerical experiments}

We conduct two numerical studies to demonstrate the robustness of
the $\gamma$-ICA procedure. In the first study, the data is
generated from independent sources with some known distributions. In
the second study, we use transformations of Lena images to form
mixed image.

\subsection{Simulated data}

\begin{figure}[h]
\centering
\includegraphics[height=6cm]{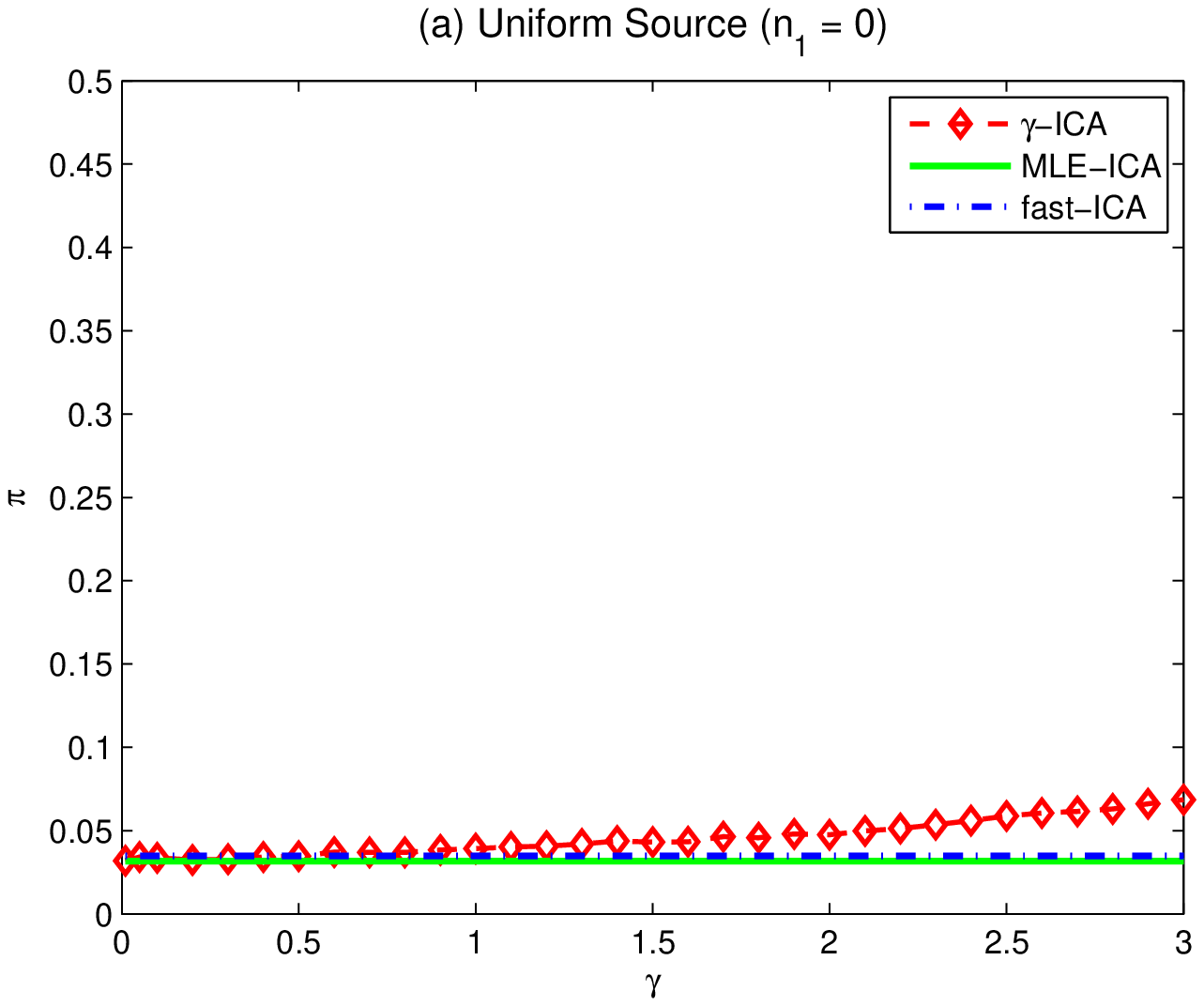}
\includegraphics[height=6cm]{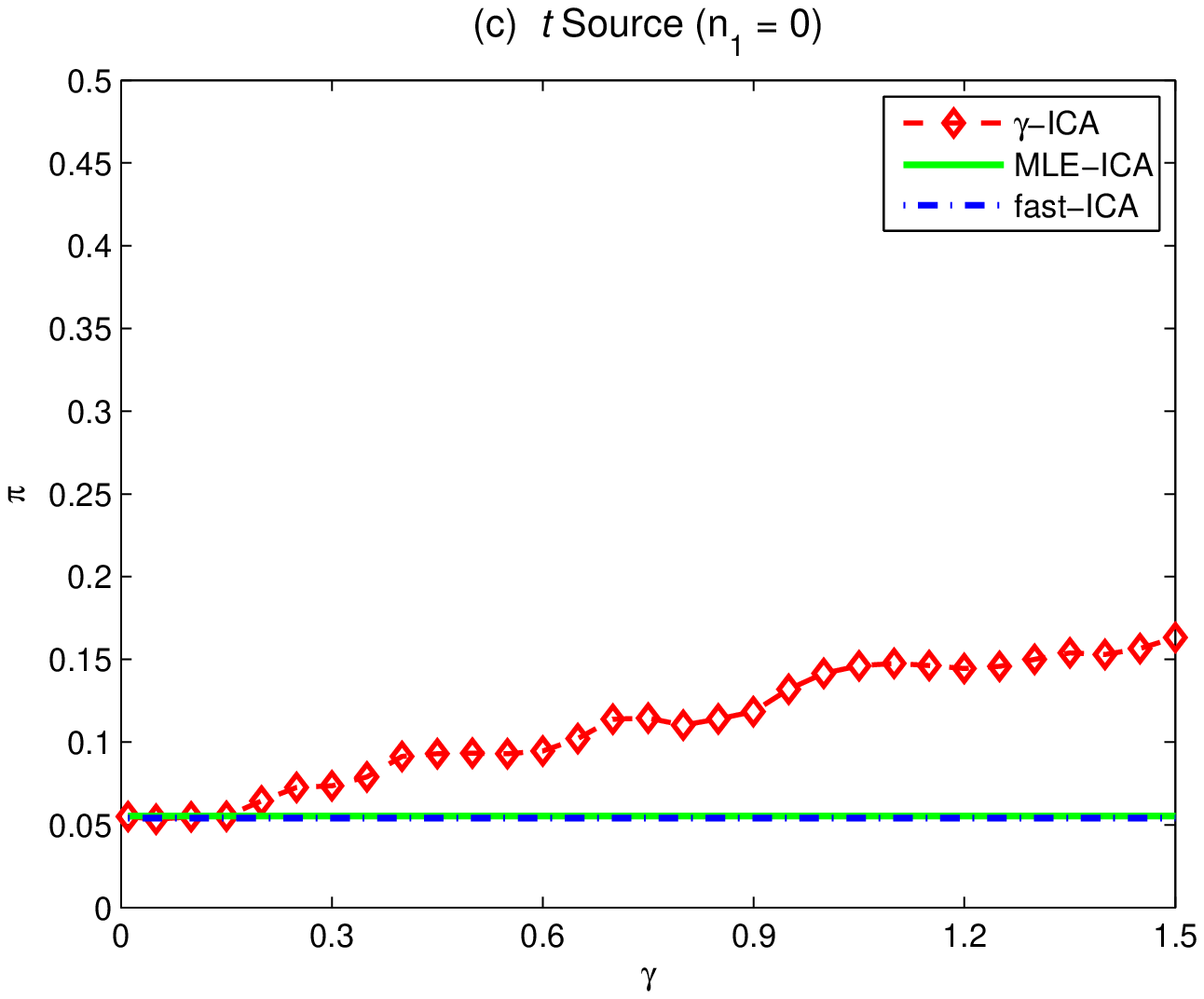}

\includegraphics[height=6cm]{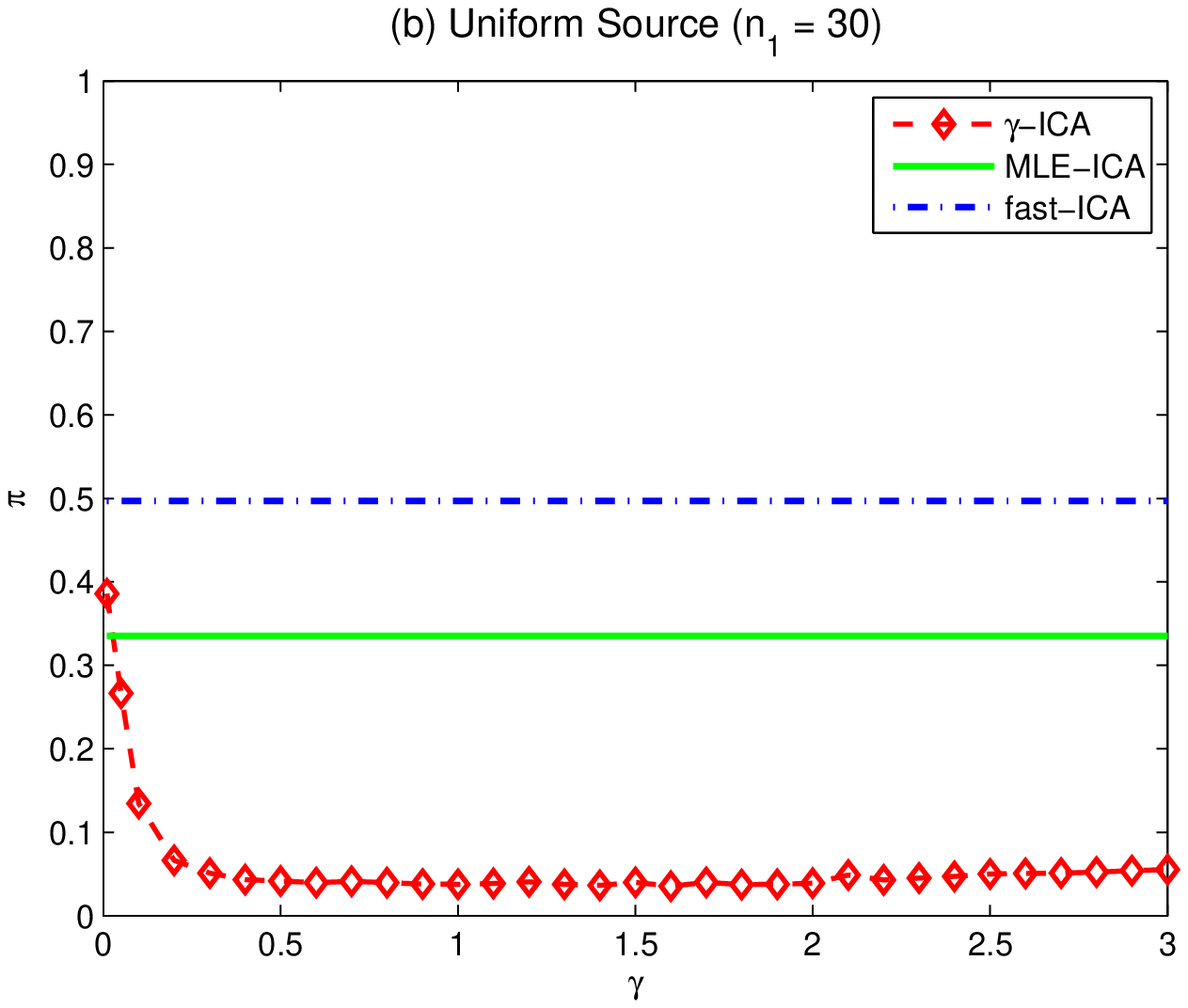}
\includegraphics[height=6cm]{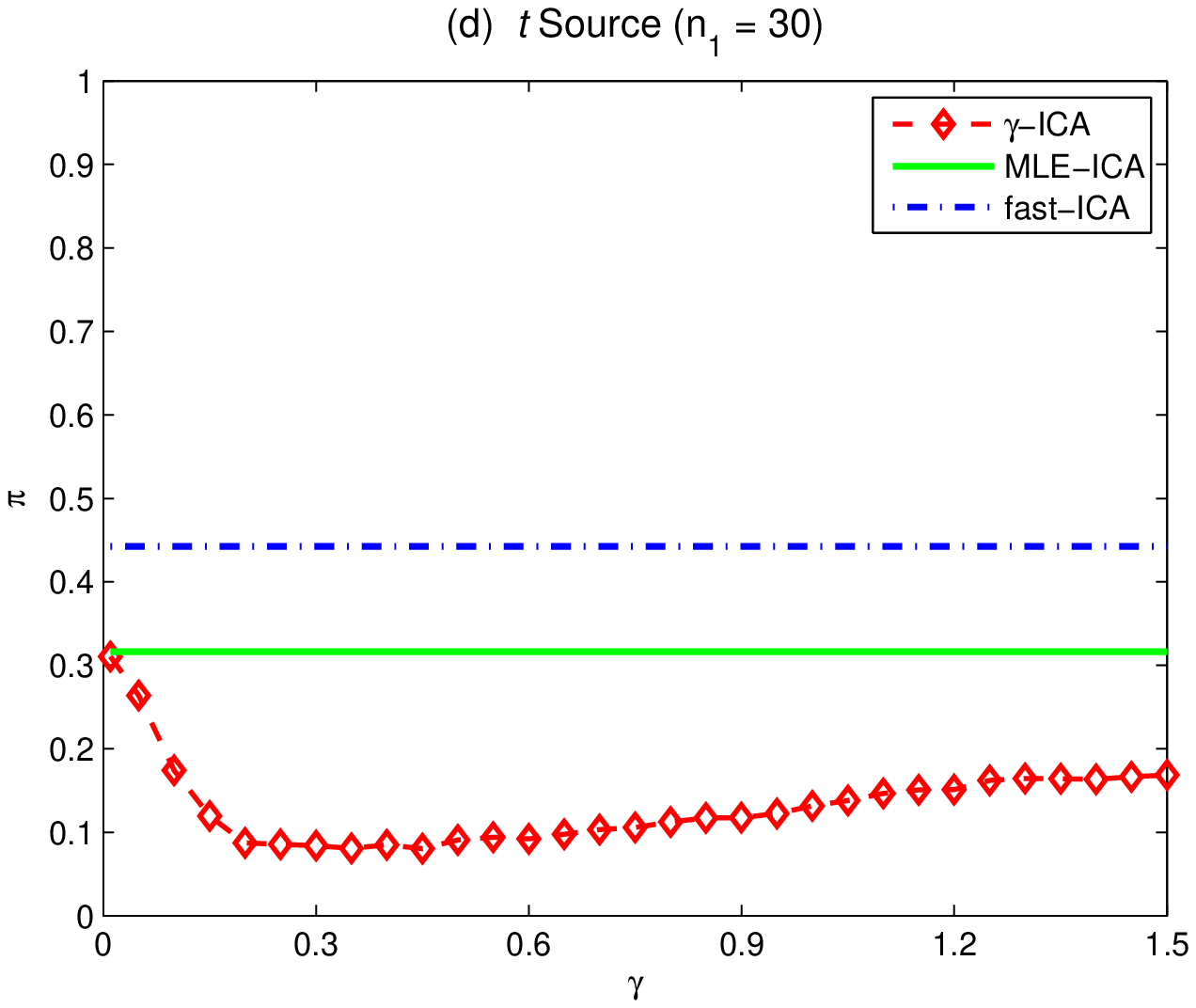}
\caption{The averages of the performance index $\pi$ for different
methods.}\label{fig.sim}
\end{figure}

We independently generate the two sources $S_j$, $j=1,2$, from a
non-Gaussian distribution with sample size $n=150+n_1$. The
observable $X$ is then given by $X=AS$, where
\begin{eqnarray*}
A=\left[\begin{array}{cc}1&2\\1&0.5
\end{array}\right].
\end{eqnarray*}
Among the $n$ observations, we add to each of the last $n_1$
observations a random noise $e$. The data thus contains $150$
uncontaminated i.i.d. observations from the ICA model, $X=AS$, and
$n_1$ contaminated i.i.d. observations from $X=AS+ e$, where $e\sim
N(\mu,\sigma^2 I_2)$ with $\mu=(5,5)$ and $\sigma=5$. We consider
two situations for the independent source $S=(S_1,S_2)$:
\begin{enumerate}
\item[(i)]
{\sc Uniform source:} Each $S_j$, $j=1,2$, is generated from
Uniform$(-3,3)$.

\item[(ii)]
{\sc Student-$t$ source:} Each $S_j$, $j=1,2$,  is generated from
$t$-distribution with 3 degrees of freedom.
\end{enumerate}
For the case of uniform source, we use the sub-Gaussian model
$f_j(s)\propto \exp(-cs^4)$ with $c=0.1$, which ensures the variance
under $f_j$ is close to unity. As to the case of $t$ source, the
super-Gaussian model $f_j(s)\propto 1/{\rm cosh}(cs)$ is considered,
and we follow the suggested range of Hyv\"arinen and Oja (2000) and
set $c=1.5$. We also implement MLE-ICA (using the geometrical
algorithm introduced in Section~4) and fast-ICA (using the code
available at \verb"http://www.cis.hut.fi/projects/ica/fastica/")
based on the $\gamma$-prewhitened data for fair comparisons. To
evaluate the performance of each method, we modify from the
performance index of Parmar and Unhelkar (2009) by a rescaling and
by replacing the 2-norm with 1-norm and define the following
performance index
\begin{eqnarray}
\pi=\frac{1}{2p(p-1)}\sum_{i}\left\{\left(\frac{\sum_{k}|\pi_{ik}|}{\max_j|\pi_{ij}|}-1\right)+
\left(\frac{\sum_{k}|\pi_{ki}|}{\max_j|\pi_{ji}|}-1\right)\right\}\le
1,
\end{eqnarray}
where $\pi_{ij}$ is the $(i,j)$-th element of $\Pi=\widetilde A\,
\widehat W^\top$. We will expect $\Pi$ to be a permutation matrix,
when the method performs well. In that situation, the value of $\pi$
should be very close to 0, and attains $0$ if $\Pi$ is indeed a
permutation matrix. Simulation results with 100 replications are
reported in Figure~\ref{fig.sim}.

\begin{figure}[h]
\centering
\includegraphics[width=\textwidth]{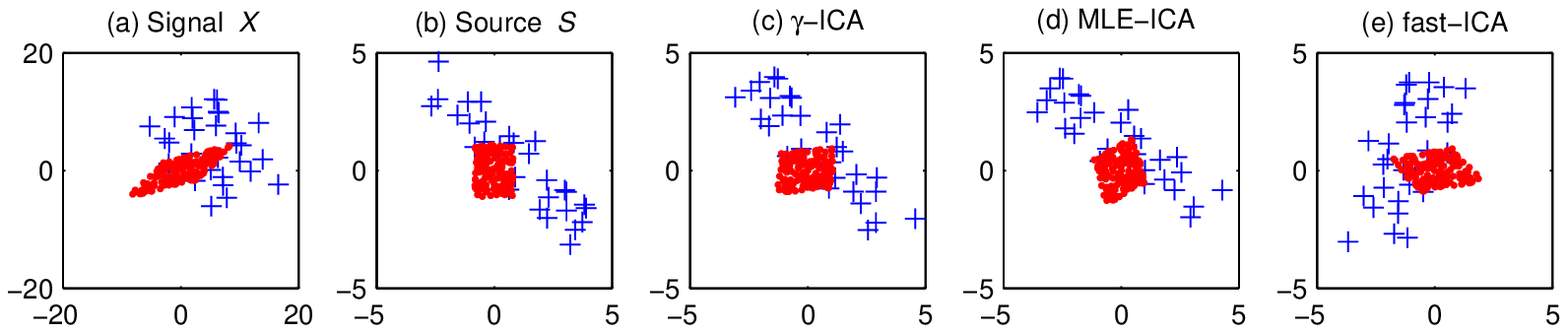}
\includegraphics[width=\textwidth]{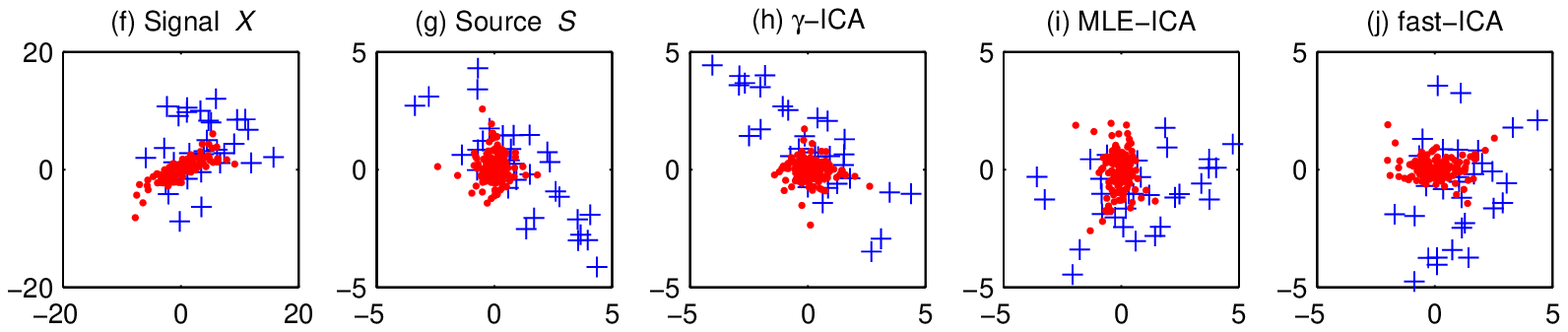}
\caption{The scatter plots for the recovered independent components
from different methods, for the observed signals $X$, and for the
true sources $S$. In each plot, the red dots are observations
without contamination, and the blue pluses are contaminated ones.
(a)-(e): Uniform source (Scenario-1), (f)-(j): $t$ source
(Scenario-2). }\label{fig.sim.scatter}
\end{figure}

For the case of no outliers ($n_1=0$), all three methods perform
well except the performance index $\pi$ of $\gamma$-ICA increases as
$\gamma$ increases. This is reasonable since, according to
Theorem~\ref{thm.consistency}, $\gamma$-ICA may fail to apply when
$\gamma$ is too large. However, this influence is not severe as the
the performance index $\pi$ is slightly increased only. As to the
case of involving outliers ($n_1=30$), it can be seen that the
proposed $\gamma$-prewhitening followed by $\gamma$-ICA does possess
the advantage of robustness for a wide range of $\gamma$ values,
while the other two methods are not able to recover the latent
sources. The performance of $\gamma$-ICA becomes worse when $\gamma$
is small, since in the limiting case $\gamma\to 0$, $\gamma$-ICA
reduces to the non-robust MLE-ICA. We note that both
$\gamma$-prewhitening and $\gamma$-ICA are critical. This can be
seen from the poor performance of MLE-ICA and fast-ICA, even they
use the $\gamma$-prewhitened data as the input. Indeed,
$\gamma$-prewhitening only ensures that we shift and rotate the data
in a robust manner, while the outliers will still enter into the
subsequent estimation process and, hence, a non-robust result is
expected. In Figure~\ref{fig.sim.scatter} we report the scatter
plots of the recovered sources $\widehat A^{-1}X$ from each method,
of $X$, and of $A^{-1}X$ for one simulation run ($n_1=30$). These
plots still convey the same message that $\gamma$-ICA is the winner
among three methods, where the pattern of the reconstructed sources
from $\gamma$-ICA is the most close to that of $A^{-1}X$.

\subsection{Lena image}

\begin{figure}[h]
\centering \hspace{0cm}\includegraphics[height=6.5cm]{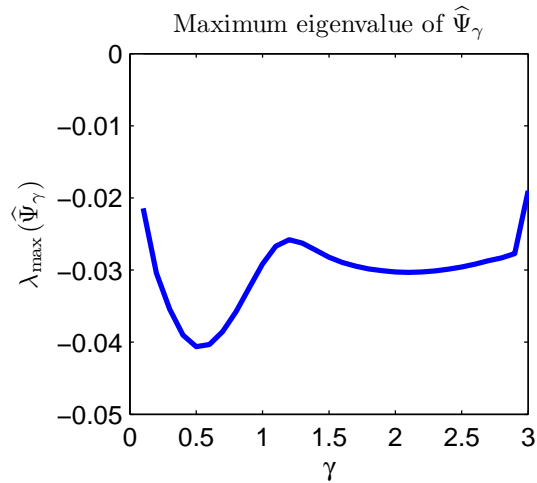}
\caption{The maximum eigenvalue of $\widehat\Psi_\gamma$ in
(\ref{hessian_tangent}) at different $\gamma$ values for the Lena
data analysis.}\label{fig.eig}
\end{figure}

\begin{figure}[h]
\centering
\includegraphics[height=6cm]{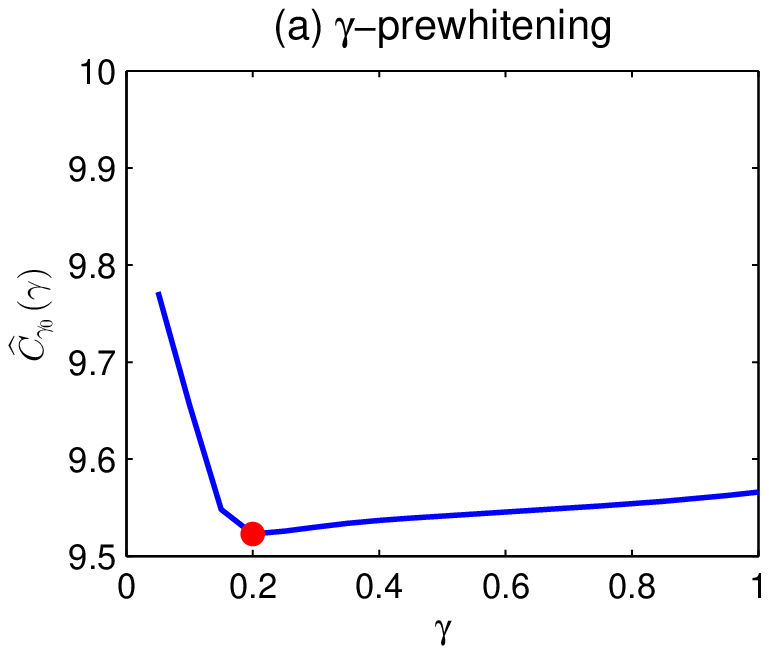}
\includegraphics[height=6cm]{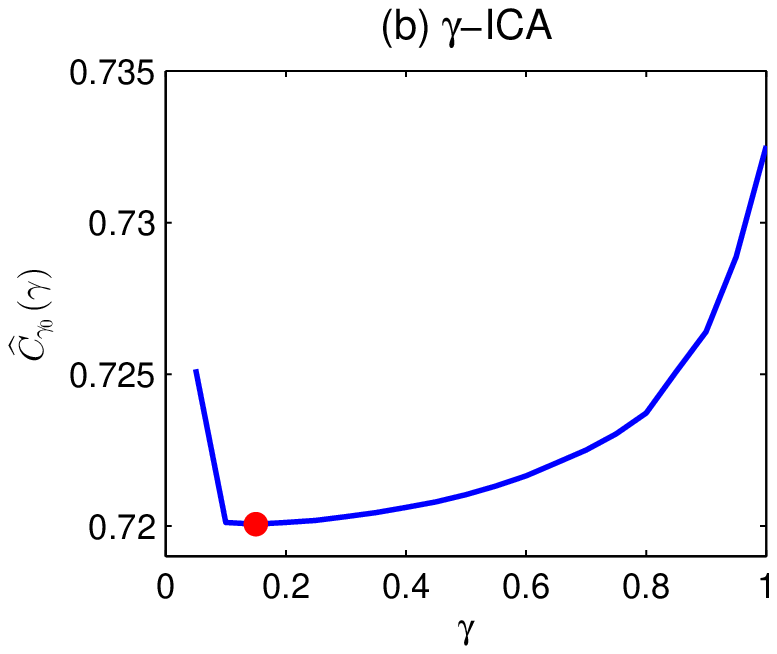}
\caption{The cross-validation estimates $\widehat
C_{\gamma_0}(\gamma)$ with $\gamma_0=1$ for (a)
$\gamma$-prewhitening and (b) $\gamma$-ICA for the Lena data
analysis. The red dot indicates the place where the minimum value
is attained.}\label{fig.cv}
\end{figure}

\begin{figure}[h]
\hspace{-1.1cm}\includegraphics[width=18cm]{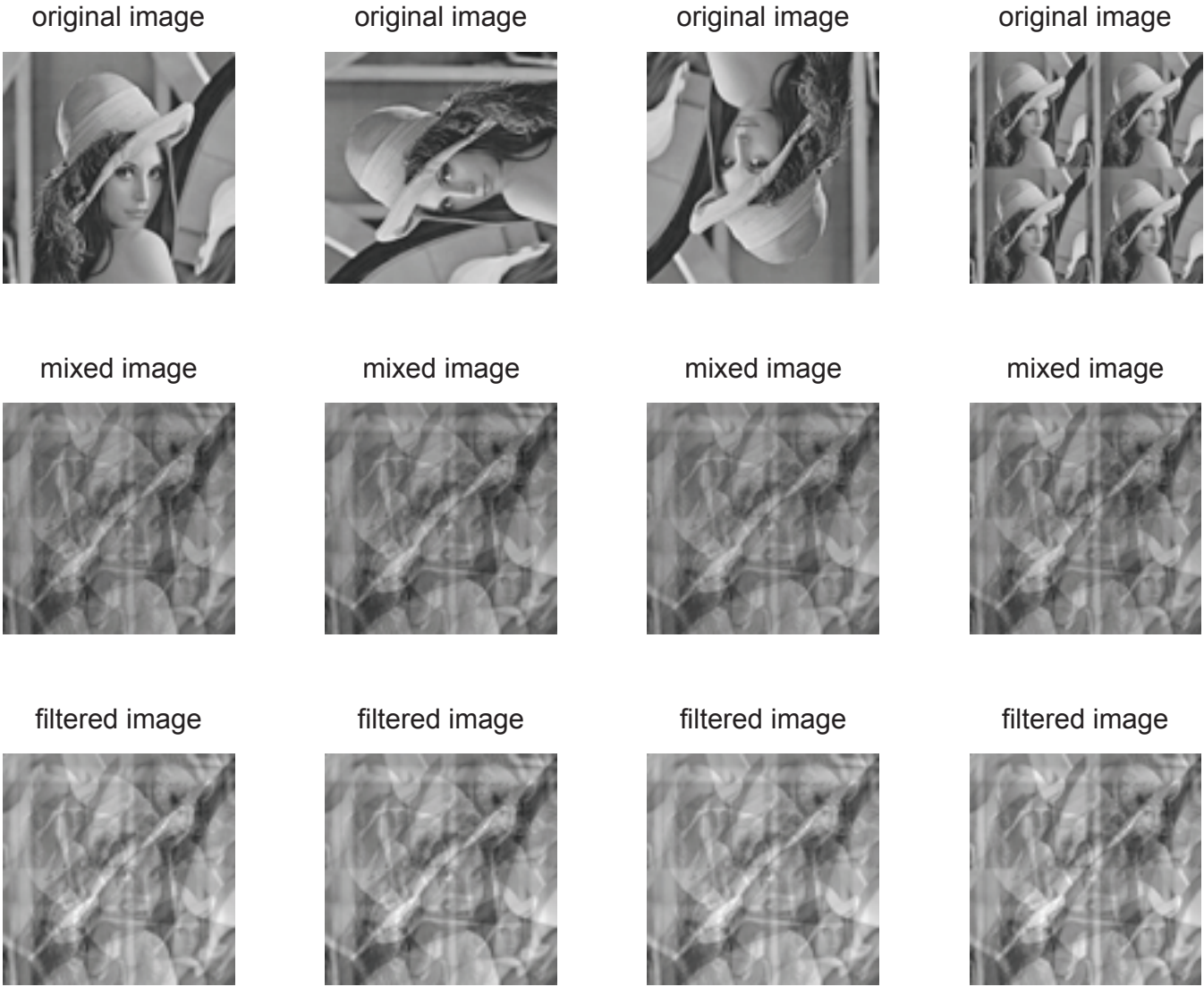} \caption{Four
images of Lena (the first row), the mixed images with $30\%$ pixels
being contaminated (the second row), and the filtered images from
the mixed images (the third row).}\label{fig.lena}
\end{figure}

\begin{figure}[h]
\hspace{-1.1cm}\includegraphics[width=18cm]{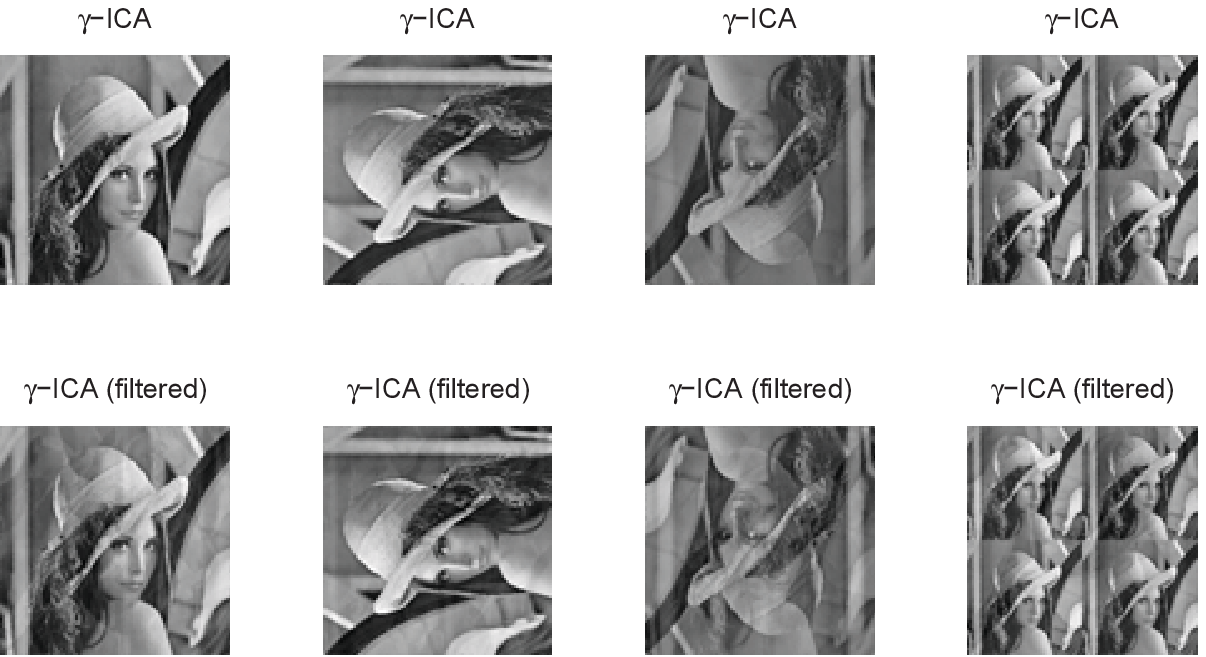}
\caption{Recovered Lena images from $\gamma$-ICA based on the mixed
images (the first row) and the filtered images (the second
row).}\label{fig.lena_gamma_recover}
\end{figure}

\begin{figure}[h]
\hspace{-1.1cm}\includegraphics[width=18cm]{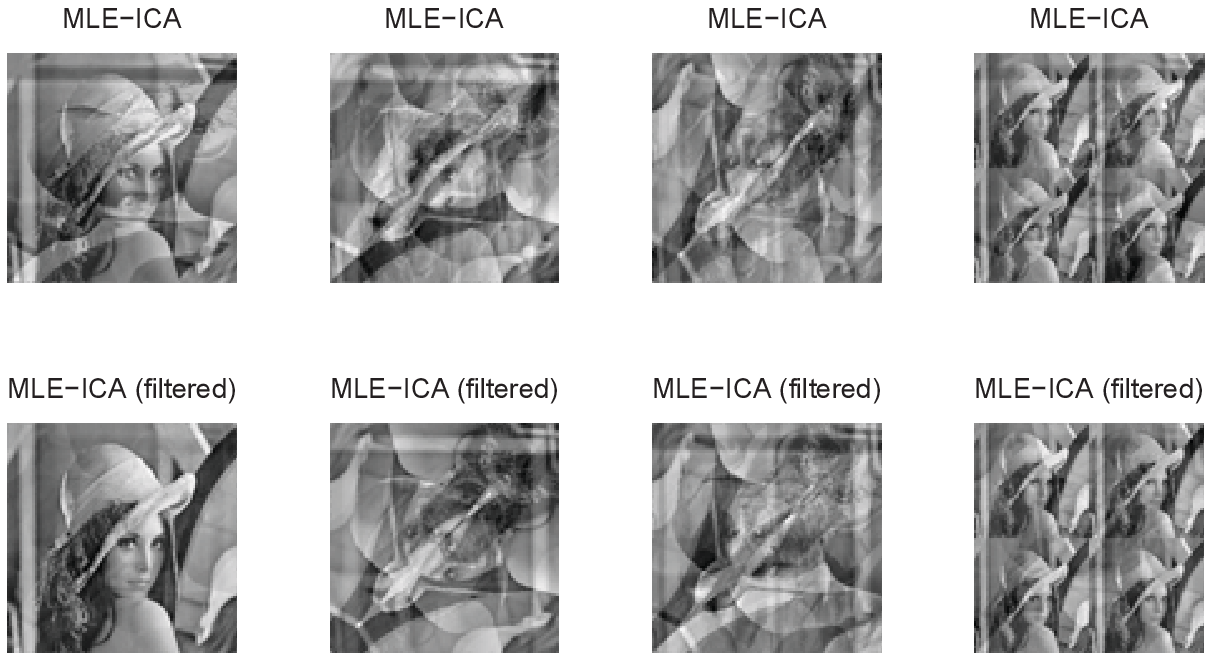}
\caption{Recovered Lena images from MLE-ICA based on the mixed
images (the first row) and the filtered images (the second
row).}\label{fig.lena_mle_recover}
\end{figure}

\begin{figure}[h]
\hspace{-1.1cm}\includegraphics[width=18cm]{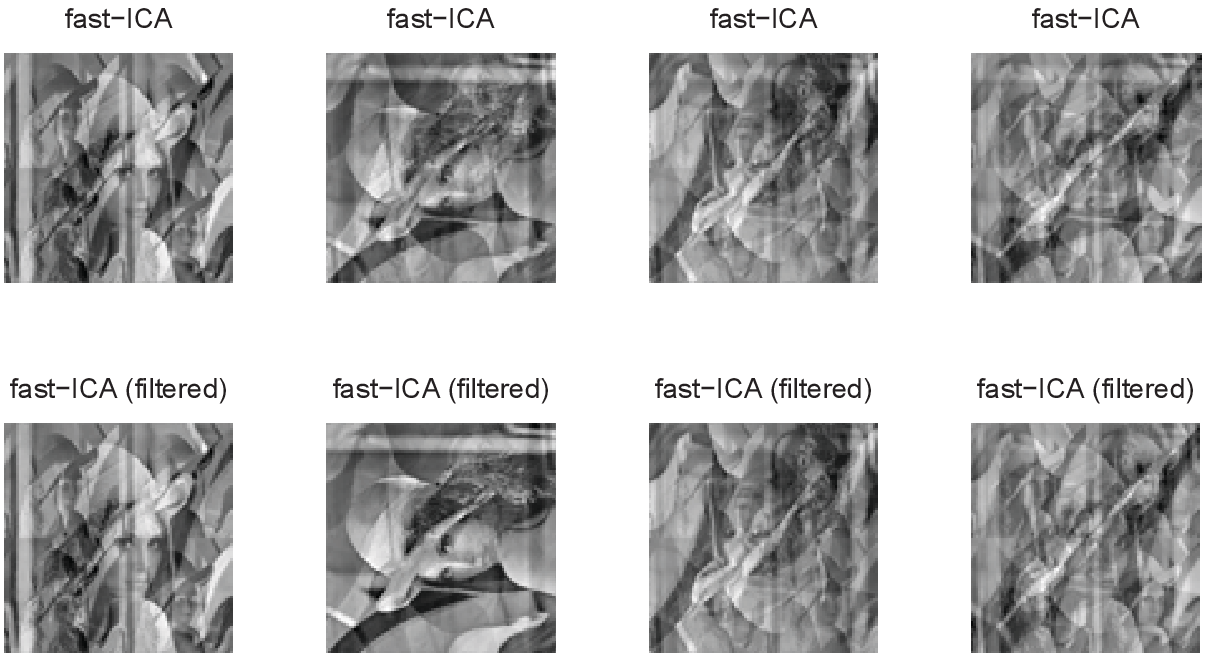}
\caption{Recovered Lena images from fast-ICA based on the mixed
images (the first row) and the filtered images (the second
row).}\label{fig.lena_fast_recover}
\end{figure}

We use the Lena picture to evaluate the performance of $\gamma$-ICA.
In our experiment, we use the Lena image with $512\times 512$
pixels. We construct four types of Lena as the latent independent
sources $S$ as shown in Figure~\ref{fig.lena}. We randomly generate
the mixing matrix to be $A=1_4 1_4^T+C$, where the elements of $C\in
\mathbb{R}^{4\times 4}$ are independently generated from
Uniform$(-0.3,0.3)$. The observed mixed pictures are also placed in
Figure~\ref{fig.lena}, wherein about $30\%$ of the pixels are added
with random noise generated from $N(20,50^2)$ for contamination. The
aim of this data analysis is to recover the original Lena pictures
based on the observed contaminated mixed pictures. In this analysis,
the pixels are treated as the random sample, each with dimension 4.
We randomly select $1000$ pixels to estimate the demixing matrix,
and then apply it to reconstruct the whole source pictures. We
conduct two scenarios to evaluate the robustness of each method:
\begin{enumerate}
\item
Using the original image $X$ as the input (see the second row of
Figure~\ref{fig.lena}).

\item
Using the filtered image $X^*$ from $X$ as the input (see the third
row of Figure~\ref{fig.lena}).
\end{enumerate}
The filtering process in the second scenario can be treated as a
pre-processing to alleviate the influence of additive Gaussian noise. In both
scenarios, the estimated demixing matrix is applied to the original
images $X$ to recover $S$. Note that with Gaussian noise contamination,
conventional prewhitening by empirical moment estimators is not
robust and, hence, both fast-ICA and MLE-ICA may fail to apply.
Therefore, we prewhiten the data by $\gamma$-prewhitening first and
then apply $\gamma$-ICA, MLE-ICA, and fast-ICA to the same whitened data for fair comparison. The plot $\{(\gamma, \lambda_{\max}(\widehat\Psi_\gamma))\}$ introduced
in the end of Section~3.3 is placed in Figure~\ref{fig.eig}, which
suggests that $\Gamma=(0,1]$ is a good candidate for possible
$\gamma$ values. We then apply the cross-validation method developed
in Section~5 to determine the optimal $\gamma\in \Gamma$. The
estimated values of $\widehat C_{\gamma_0}(\gamma)$ are plotted in
Figure~\ref{fig.cv}, from which we select
$\widehat\gamma_{\mu,\Sigma}=0.2$ for $\gamma$-prewhitening and
$\widehat\gamma_W=0.15$ for $\gamma$-ICA. The recovered pictures are
placed in
Figures~\ref{fig.lena_gamma_recover}-\ref{fig.lena_fast_recover},
where for each figure the first row is for Scenario-1 and the second
row is for Scenario-2.

It can be seen that $\gamma$-ICA is the best performer under both
scenarios, while MLE-ICA and fast-ICA cannot recover the source
images well when data is contaminated. It also demonstrates the
applicability of the proposed $\gamma$-selection procedure. We
detect that MLE-ICA and fast-ICA perform better when using filtered
images $X^*$, but can still not reconstruct images as good as
$\gamma$-ICA does. Interestingly, $\gamma$-ICA has a reverse
performance, where the best reconstructed images are from the
original images instead of the filtered ones. The filtering process,
which aims to achieve robustness, replaces the original pixel value
by the median of the pixel values over its neighborhood. Therefore,
while filtering process will alleviate the influence of outlier, it
is also possible to lose useful information at the same time. For
instance, a pixel without being contaminated will still be replaced
by certain median value during the filtering process. $\gamma$-ICA,
however, works on the original data $X$ that possesses all the
information available, and then weights each pixel according to its
observed value to achieve robustness. Hence, a better performance
for $\gamma$-ICA based on the original images is reasonably
expected.

\section{Conclusions}

In this paper, we introduce a unified estimation framework by means
of minimum $U$-divergence. For the reason of robustness
consideration, we further focus on the specific choice of
$\gamma$-divergence, which gives the proposed $\gamma$-ICA inference
procedure. Statistical properties are rigorously investigated. A
geometrical algorithm based on gradient flows on orthogonal group is
introduced to implement our $\gamma$-ICA. The performance of
$\gamma$-ICA is evaluated through synthetic and real data examples.

There are still many important issues that are not covered by this
work. For example, we only consider full ICA problem, i.e., simultaneous extraction of all $p$
independent components, which is unpractical in the
case of large $p$. It is of interest to extend our current $\gamma$-ICA to
partial $\gamma$-ICA. Another issue of interest is also related to the large-$p$-small-$n$
scenario. In this work, data have to be prewhitened before entering the
$\gamma$-ICA procedure. Prewhitening can be very unstable
especially when $p$ is large. How to avoid such a difficulty is an interesting and challenging issue. Tensor data analysis is now becoming popular and attracts the attention of many
researchers. Many statistical methods include ICA have been extended
to deal with such a data structure by means of multilinear algebra
techniques. An extension of $\gamma$-ICA to a multilinear setting to cover tensor data analysis is also of great interest for future study.


\begin{center}
\textbf{\Large References}
\end{center}

\begin{description}
\item
Amari, S., Chen, T. and Cichocki, A. (1997). Stability analysis of
learning algorithms for blind source separation. {\it Neural
Networks}, 10, 1345-1351.

\item
Amari, S. (1998).
Natural gradient works efficiently in learning.
{\it Neural Computation}, 10, 251-276.
\item
Basu, A., Harris, I. R., Hjort, N. L. and Jones, M. C. (1998).
Robust and efficient estimation by minimizing a density power
divergence. {\it Biometrika} 85, 549-559.

\item
Bertsekas, D. P.  (2003). {\it Nonlinear Programming}.
 Athena Scientific, Belmont, Massachusetts.

\item
Boothby, W. M. (1986). {\it An Introduction to Differentiable
Manifolds and Riemannian Geometry}. Academic Press.

\item
Edelman, A., Aris, T. A. and Smith, S. (1998). The geometry of
algorithm with orthogonality constraints {\it SIAM J. Matrix Anal.
Appl.} 20,  303-353.

\item
Eguchi, S. (2009). Information divergence geometry and the
application to statistical machine learning. In {\it  Information
Theory and Statistical Learning}, F. Emmert-Streib and M. Dehmer (eds.), 309-332.
Springer, Berlin.

\item
Fiori, S. (2005). Quasi-geodesic neural learning algorithm over the
orthogonal group: a tutorial. {\it J. Machine Learning Research}, 6,
743-781.

\item
Fujisawa, H. and Eguchi, S. (2008). Robust parameter estimation with
a small bias against heavy contamination. {\it Journal of
Multivariate Analysis}, 99, 2053-2081.


\item
Horn, R. A. and Johnson,  C. R. (1991). {\it Topics in Matrix
Analysis}. Cambridge University Press, Cambridge; New York.

\item
Hyv\"arinen, A. (1998). New approximations of differential entropy
for independent component analysis and projection pursuit. {\it
Advances in Neural Information Processing Systems}, 10, 273-279.

\item
Hyv\"arinen, A. (1999). Fast and robust fixed-point algorithms for
independent component analysis. {\it IEEE Transactions on Neural
Networks}, 10, 626-634.

\item
Hyv\"arinen, A. and Oja, E. (2000). Independent component analysis:
algorithm and applications. {\it Neural Networks}, 13, 411-430.

\item
Hyv\"arinen, A., Karhnen, J. and Oja, E. (2001). {\it Independent
Component Analysis.} Wiley Inter-Science.

\item
Magnus, J. R. and Neudecker, H. (1979). The commutation matrix: some
properties and applications. {\it Annals of Statistics}, 7,
381--394.

\item
Marsden, J. E. and Ratiu, S. T. (1998).
{\it Introduction to Mechanics and Symmetry: A Basic Exposition of Classical Mechanical Systems}. Springer.

\item
Mihoko, M. and Eguchi, S. (2002). Robust blind source separation by
$\beta$-divergence. {\it Neural Computation}, 14, 1859-1886.

\item
Minami, M. and Eguchi, S. (2003). Adaptive selection for minimum
$\beta$-divergence method. \textit{Proceedings of ICA-2003
Conference}, Nara, Japan.

\item
Mollah, M. N. H., Eguchi, S. and Minami, M. (2007). Robust
prewhitening for ICA by minimizing $\beta$-divergence and its
application to fastICA. {\it J. Neural Processing Letters}, 25,
91-110.

\item
Murata, N., Takenouchi, T., Kanamori, T. and Eguchi, S. (2004).
Information geometry of U-boost and Bregman divergence. {\it Neural
Computation}, 16, 1437-1481.

\item
Nishimori, Y. and Akaho, S. (2005). Learning algorithms utilizing
quasi-geodesic flows on the Stiefel manifold. {\it Neurocomputing},
67, 106-135.

\item
Parmar, S. D. and Unhelkar, B. (2009). Performance analysis of ICA
algorithms against multiple-sources interference in biomedical
systems. {\it International Journal of Recent Trends in
Engineering}, 2, 19-21.
\end{description}

\newpage

\end{document}